\documentclass{iau} 
\usepackage{graphicx}

\title[Comparison of Low-Mass and High-Mass Star Formation] 
{Comparison of Low-Mass and High-Mass Star Formation}

\author[Jonathan C. Tan]   
{Jonathan C. Tan$^1$}

\affiliation{$^1$Depts. of Astronomy \& Physics, University of Florida, Gainesville, FL 32611, USA\\ email: {\tt jctan.astro@gmail.com}}

\pubyear{2015}
\volume{315}  
\jname{From interstellar clouds to star-forming galaxies: universal processes?}
\editors{P. Jablonka, P. Andr\'e, F. F. S. van der Tak, eds.}
\begin{document}

\maketitle

\begin{abstract}
I review theoretical models of star formation and how they apply
across the stellar mass spectrum. Several distinct theories are under
active study for massive star formation, especially {\it Turbulent
  Core Accretion}, {\it Competitive Accretion} and {\it Protostellar
  Mergers}, leading to distinct observational predictions. These
include the types of initial conditions, the structure of infall
envelopes, disks and outflows, and the relation of massive star
formation to star cluster formation. Even for Core Accretion models,
there are several major uncertainties related to the timescale of
collapse, the relative importance of different processes for
preventing fragmentation in massive cores, and the nature of disks and
outflows. I end by discussing some recent observational results that
are helping to improve our understanding of these processes.
%
\end{abstract}

\firstsection 
\section{Introduction}

A universal theory of star formation that successfully describes the
birth of low- and high-mass stars is a long-sought goal. In
particular, large efforts have been made to see if massive stars form
in a similar way to low-mass stars, i.e., via the {\it Core Accretion}
paradigm, starting with pre-stellar cores (PSCs) that then collapse to
form a single or binary protostar, with most matter accreting via a
rotationally-supported disk that also helps to launch protostellar
outflows (e.g., Shu et al. 1987; Inutsuka 2012). However, alternatives
to this mechanism include {\it Competitive Accretion} (Bonnell et
al. 2001; Wang et al. 2010) and {\it Protostellar Mergers} (Bonnell et
al. 1998; Bally \& Zinnecker 2005; Moeckel \& Clarke 2011). Both have
been proposed for massive star formation in crowded regions of
protostellar clusters, the most common, perhaps near-universal, sites
where massive stars form. However, these regions are also relevant for
the bulk of star formation in galaxies and, in fact, Competitive
Accretion has been invoked to explain most of the range of the stellar initial
mass function (IMF) $\gtrsim 1\:M_\odot$ (e.g., Bate
2012). Here I discuss latest work that seeks to elucidate the
formation mechanism of massive stars, including expected similarities
and differences with low-mass protostars and role of environment.


\section{Massive Star Formation Environments}

Massive stars have dominated the universe since its earliest
epochs. The first, ``Pop III'' stars, forming in individual dark
matter minihalos from pure H \& He gas, are thought to have been
$\gtrsim 10-100\:M_\odot$, so that they then initiated reionization
and chemical enrichment of their surroundings (e.g., Bromm
2013). Continued radiative, mechanical and chemical feedback from
massive stars is then theorized to have shaped the formation and
evolution of galaxies (e.g., Vogelsberger et
al. 2014). Observationally, the light seen from distant galaxies is
dominated by that from massive stars (e.g., Sobral et al. 2015). This
is true also in regions of nearby galaxies with active star formation,
where the clustered nature of the process is readily apparent (e.g.,
Whitmore et al. 2014). There is a wide range of mass surface densities
of these young star clusters from $\Sigma\sim0.1$--$30\:{\rm
  g\:cm^{-2}}$ and a narrower range of sizes of $\sim1$--10~pc (Tan et
al. 2014 [T14]). Driven by these star clusters and dissolving OB
associations, feedback on the interstellar medium is powered via
radiation pressure, stellar winds, ionization and supernovae. Such
feedback is likely to play a prominent role in the regulation of star
formation rates (e.g., Walch et al. 2015).

Large fractions of all stars, including massive stars, form in
clusters (e.g., de Wit et al. 2005; Gutermuth et al. 2009) and the
distribution of star formation is much more clustered than that of
molecular gas. This means that most stars and planetary systems may
have been strongly influenced by massive star feedback at the time of
their birth.

Another environment where massive stars form is galactic center
regions, including in dense circumnuclear starburst disks, where mean
disk mass surface densities can range from $\Sigma \sim
0.1$--$100\:{\rm g\:cm^{-2}}$ (Downes \& Solomon 1998; Wilson et
al. 2014). Our Galactic center has a large population of young
($\lesssim 6$~Myr), relatively massive stars, some orbiting in a
disk-like structure around the supermassive black hole (e.g., Lu et
al. 2014). The growth of supermassive black holes may be regulated by
such star formation activity.

\section{The Physics of Star Formation}\label{S:physics}

The physics of star formation involves a competition between
self-gravity of gas clumps (i.e., structures that fragment into star
clusters) and cores (i.e., structures that collapse to a central
rotationally-supported disk that forms a single star or small-$N$
multiple) and processes that resist collapse, i.e., various kinds of
pressure forces, such as thermal, turbulent and magnetic. The
evolution of these pressures needs to be followed by considering
heating and cooling processes, generation and decay of turbulence, and
generation and diffusion of $B$-fields. Once the clump or core
contains stars, then support against collapse may be provided by
mechanical feedback (i.e., protostellar outflows or stellar winds),
radiation pressure (acting mostly on dust), or enhanced thermal
pressure from photoionized regions. The chemical evolution of the gas
and dust needs to be followed during collapse, especially as this
helps set the trace ionization fraction that is important for coupling
$B$-fields to the mostly neutral gas. Rotational support becomes
important in protostellar disks, but theoretical prediction of their
scale is quite uncertain due to the dominant role of magnetic braking in
transferring angular momentum (Li et al. 2014). Fragmentation of gas
in the clump and in the core's disk will depend on the evolution of
the local pressure and rotation/shear support contributions.  To
accurately model star formation, a wide range of spatial and temporal
scales extending down to those of the protostellar surface need to be
followed. Unlike the cosmological Pop III case, numerical simulations
of ``local'' star formation face the additional challenge of uncertain
choices for initial conditions: e.g., how close is the initial
clump to virial and pressure equilibrium?

Due to the complexity of this highly nonlinear and high dynamic range
problem, which remains beyond full computational tractability, there
are many open questions. What causes a certain region of molecular
cloud to form stars, i.e., does it typically occur due to an external
trigger (e.g., converging flows, cloud collisions or stellar feedback)
or via spontaneous gravitationally instability (e.g., as a cloud
evolves and loses its earlier level of internal pressure support)?
This question also relates to the specification of the initial
conditions. Then, what is the accretion mechanism of the protostar,
i.e., is most of the mass already organized into a self-gravitating
PSC or is it later accreted competitively to the protostar from a
previously unbound state from the larger-scale clump?  Are
protostellar mergers important? Are protostellar interactions that may
disturb cores and disks important?  What is the timescale of
individual star formation: e.g., does a core collapse on a timescale
that is similar to or much slower than the local free-fall time (for
convenience defined with reference to that of a uniform density
sphere: $t_{\rm ff}\equiv(3\pi/[32G\rho])^{1/2}$)?  Similarly, is the
clump undergoing rapid global collapse to form a star cluster
(Elmegreen 2000; Hartmann \& Burkert 2007) or is the process much
slower allowing the clump to be in quasi equilibrium (Tan et
al. 2006), perhaps regulated by outflow-driven turbulence (Nakamura \&
Li 2014)? A consequence of these different possibilities is the range
of age spreads present in young star clusters, again relative to the
local clump free-fall time (Da Rio et al. 2014). Are massive stars the
first to start forming in a cluster, becoming massive because of this
head start (Wang et al. 2010; Bate 2012), or are their individual
formation times short compared to the duration of cluster formation
(McKee \& Tan 2002), allowing contemporaneous formation with
lower-mass stars that is potentially independent of the evolutionary
stage of the protocluster? What processes shape the stellar IMF and
initial binary properties? How do these quantities vary with
environment?

\subsection{Turbulent Core Accretion}

A variety of Core Accretion models for massive star formation have
been proposed, extending the basic concepts of low-mass star formation
models (see T14 for a review). The Turbulent Core Accretion model
(McKee \& Tan 2003 [MT03]) recognizes that thermal pressure is
unimportant for supporting massive PSCs (i.e., they are much more
massive than the local Jeans or Bonnor-Ebert mass), so such support
must come from some combination of nonthermal means, i.e., turbulence
and/or magnetic fields. The next key assumption is that the pressure
of the clump environment, which is $P_{\rm cl} \simeq G \Sigma_{\rm
  cl}^2$ for a self-gravitating cloud, sets the surface boundary
condition of the core of a given mass. PSCs are modeled as singular
polytropic spheres not too far from virial and hydrostatic
equilibrium. Then the radius is given by $R_c = 0.057
M_{c,60}^{1/2}\Sigma_{\rm cl,1}^{-1/2}$~pc, where $M_{c,60}$ is the
core mass normalized by $60\:M_\odot$ and $\Sigma_{\rm cl,1}$ is the
clump mass surface density normalized by $1\:{\rm g\:cm^{-2}}$. This
is the degree to which the core mass must be concentrated in order to
become gravitationally unstable. Note that no assumption has been made
about whether the PSC has taken a short or long time to form from the
clump, although achieving approximate pressure and virial equilibrium
is expected to take at least $\sim 1 t_{\rm ff}$. In principle, it is
possible to estimate the age of a given PSC via comparison of
predictions of astrochemical models, especially of deuterated species
(e.g., Kong et al. 2015a), with observed abundances
(\S\ref{S:PSC}). MT03 assumed cores and clumps have an internal radial
density gradient $\rho\propto r^{-k_\rho}$, with $k_\rho\simeq 1.5$
set empirically (and also consistent with later studies of Infrared
Dark Clouds (IRDCs) clumps/cores; Butler \& Tan 2012 [BT12]).

The core is then assumed to undergo inside-out collapse at a rate
comparable with that of local free-fall collapse, i.e., $\dot{m}_{*d}
= 9.3\times 10^{-4} \epsilon_{*d} (M_{c,60}\Sigma_{\rm
  cl,1})^{3/4}(M_{*d}/M_c)^{1/2}\:M_\odot\:{\rm yr}^{-1}$, where
$\dot{m}_{*d}$ is the rate of increase of the mass of the protostar
and its disk, $\epsilon_{*d}$ is the current efficiency of the infall
rate with respect to uninhibited collapse (values $\sim 0.5$ are
expected due to protostellar outflow feedback, e.g., Zhang et al. 2014
[ZTH14]) and $M_{*d}$ is the idealized collapsed mass supplied to the
central disk in the no-feedback limit. Note, this accretion rate
estimate can be applied to stars forming from cores of all masses,
with model assumptions only beginning to break down close to the
Bonner-Ebert mass, $M_{\rm BE}=0.050 (T/20\:{\rm K})^2\Sigma_{\rm
  cl,1}^{-1}\:M_\odot$. The timescale for star formation is $t_{*f} =
1.3\times 10^{5} M_{c,60}^{1/4}\Sigma_{\rm cl,1}^{-3/4}\:{\rm yr}$,
which has a very weak dependence on core mass. This timescale is
similar to the clump's free-fall time. Thus it is important to know if
the clump is undergoing rapid, free-fall collapse, or whether star cluster
formation is a slower, more drawn-out affair.

The question of what, if anything, prevents fragmentation of massive
PSCs is crucial, since this physics may play a decisive role in
shaping the IMF via the (pre-stellar) core mass function (PS)CMF. In
the Turbulent Fragmentation model of Padoan \& Nordlund (2002), a
power law spectrum of super-Alfv\'enic turbulence is assumed and the
size of dense cores is associated with the thickness of post-shock
gas, so cores are born in a very dynamic fashion. This model has been
compared to numerical simulations by Padoan et al. (2007; see also
Collins et al. 2011; Hennebelle et al. 2011), who point out the
potential importance of $B$-fields in setting the CMF. Hennebelle \&
Chabrier (2008) presented a theory of the CMF based on the
Press-Schecter statistical formalism applied to supersonic
turbulence. In their model, more massive cores derive more support
from turbulence, but also require lower density initial conditions to
help set their larger ``turbulent Jeans mass,'' than lower-mass
cores. Alternatively, Kunz \& Mouschovias (2009) proposed a dominant
role for $B$-fields, with the CMF being set by ambipolar
diffusion. Motivated by their studies of massive, cold IRDC cores that
contain $\gtrsim 100\:M_{\rm BE}$, BT12 proposed $B$-field support was
the dominant mechanism inhibiting fragmentation, so that $M_c \simeq
M_B=38 (B_c/{\rm mG})^3 (n_{\rm H,c}/10^{6}{\rm
  cm^{-3}})^{-2}\:M_\odot$, the magnetic critical mass (Bertoldi \&
McKee 1992). One should note that massive starless cores, like massive
stars, will be rare objects within the clump, and this rarity may be
set by being in the high $B$-field-strength tail of the
distribution. Krumholz \& McKee (2008) proposed radiative heating from
surrounding lower-mass protostars prevents fragmentation of massive
PSCs, which requires high accretion luminosities and thus high
$\dot{m}_{*d}$ and so high $\Sigma_{\rm cl}$, $\gtrsim 1\:{\rm
  g\:cm^{-2}}$. This model predicts massive star formation requires
the presence of an already forming protocluster and that massive
protostars would not form from cold, dense cores within IRDCs.

Many numerical simulations of the collapse of massive, turbulent
``cores'' have been carried out. Dobbs et al. (2005) presented
hydrodynamics-only simulations, finding extensive fragmentation into
stars with masses of only $\sim0.1\:M_\odot$, and thus argued for
Competitive Accretion. Krumholz et al. (2007) included dust
reprocessed radiative feedback that heated the gas, raised the Jeans
mass and thus reduced fragmentation. These simulations also showed
radiation pressure does not prevent the formation of massive stars,
given the high optical depths of the accretion flows. The strong
influence of $B$-fields is illustrated in the results of Peters et
al. (2011), who simulated a 100$\:M_\odot$ core with a $10\:{\rm \mu
  G}$ field that suffered extensive fragmentation, and those of
Seifried et al. (2011) and Myers et al. (2013), who included $\sim$mG
$B$-fields and found very limited fragmentation.

\begin{figure}[t]
\vspace*{-0.235 cm}
\begin{center}
 \includegraphics[width=4.96in]{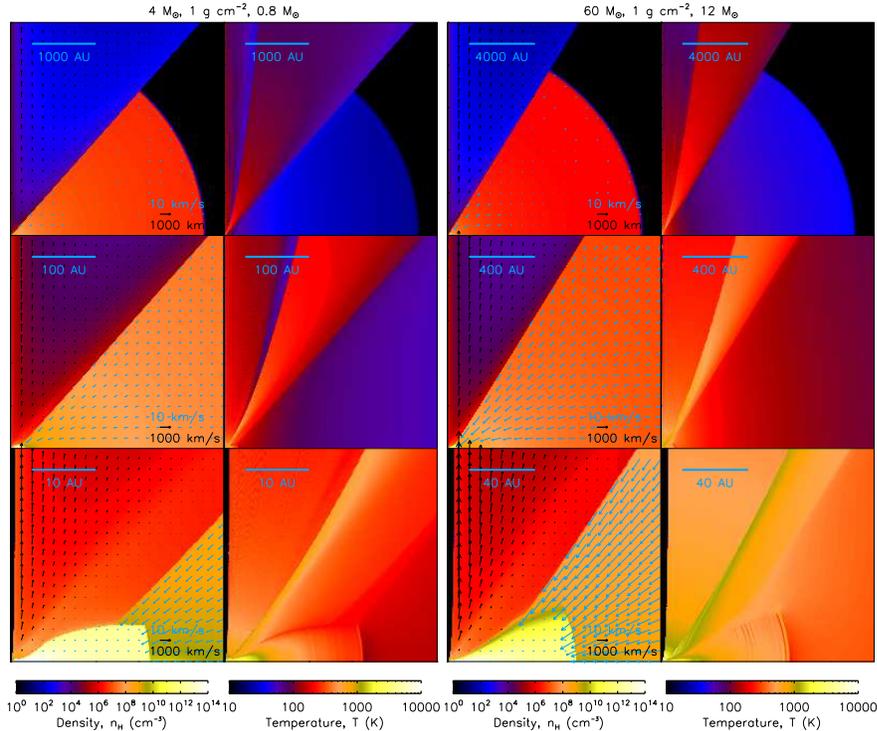} 
\vspace*{-0.45cm}
 \caption{
Comparison of low- (Zhang \& Tan 2015) and high-mass (ZTH14)
protostars forming via Turbulent Core Accretion. Left 6 panels show
slices through a $4\:M_\odot$ core in a $1\:{\rm g\:cm^{-2}}$ clump
that has collapsed to form a protostar with $0.8\:M_\odot$ (i.e., 20\%
of initial core mass), which is located in the bottom left corner of
each panel. Left column shows H nuclei number density, $n_{\rm H}$,
and inflow/outflow velocities (note different arrow scales). Right
column shows temperature. The view zooms from scale of the core (top
row), to inner infall envelope (middle), to accretion disk
(bottom). Right 6 panels show the equivalent figures for a
$60\:M_\odot$ core in the same clump environment that has collapsed to
form a $12\:M_\odot$ protostar (also 20\% of initial $M_c$). Note
linear scales in each row are $\simeq4\times$ larger than the low-mass
case. While overall morphologies and absolute densities are quite similar,
temperatures are significantly greater in the high-mass case, given
its much larger luminosity of $5\times10^4\:L_\odot$ (cf., $300\:L_\odot$
of the low-mass protostar).
}
\label{fig:protostars}
\end{center}
\end{figure}

The analytic MT03 Turbulent Core model of collapse of massive cores in
the limit of no fragmentation has been coupled to semi-analytic models
of accretion disk structure (Zhang \& Tan 2011), disk-wind
protostellar outflows (Zhang et al. 2013b) and a 1D numerical
protostellar evolution code (ZTH14). These models have been used as
inputs for continuum radiative transfer simulations to predict
multiwavelength images and spectral energy distributions
(SEDs). Lower-mass cores were studied by Zhang \& Tan (2015). Effects
of varying the main environmental variable, $\Sigma_{\rm cl}$, have
been explored.
Density and temperature structures around example low- \&
high-mass protostars are shown in Fig.~\ref{fig:protostars}.

Model predictions include the strong rise of fluxes at
$\lambda<100\:{\rm \mu m}$ as protostars grow in mass and luminosity
and open up outflow cavities. Intensity profiles along the outflow
axis are typically very asymmetric at $\lambda\lesssim 40\:{\rm \mu
  m}$, with the near-facing outflow cavity being much brighter than
the far-facing one. This asymmetry decreases for wavelengths $\gtrsim
70\:{\rm \mu m}$ at which the core envelope is becoming optically
thin. Protostellar cores in high $\Sigma_{\rm cl}$, i.e., high
pressure, environments are smaller, denser, have higher accretion
rates and luminosities, and are thus warmer, than cores of the same
mass in low $\Sigma_{\rm cl}$ regions. High-mass envelopes and disks
have similar overall morphologies and densities as their lower-mass
counterparts, but, given the more luminous central protostar, are
warmer.

In general, the advantages of the ZTH14 models for massive star
formation are that they include: detailed, self-consistent physical
models for the core structure and protostellar evolution; gas, as well
as dust, opacities; and core properties that respond to clump
environment, so typical massive cores are much smaller and denser and
collapse with higher accretion rates compared to those of other models
(e.g., Indebetouw et al. 2006; Robitaille et al. 2006). However, they
are still highly idealized in having: axisymmetric, smooth density
structures (cf., Indebetouw et al. 2006); a simplified treatment of
the interface between outflow cavity and infall envelope; and
do not treat binary protostars.


An important difference of massive protostars compared to low-mass
ones is that they can become significant sources of FUV and EUV
radiation, since their accretion timescales are generally longer than
their Kelvin-Helmholz contraction timescales. EUV photons should first
lead to ionization of the protostellar outflows, i.e., initially in a
phase of an ``Outflow-Confined HII Region'' (Tan \& McKee 2003;
Tanaka et al. 2015). Here confinement is mostly in lateral
directions, i.e., the disk and infall envelope are shielded. However,
ionization soon becomes unconfined in directions along the outflow,
which thus leads to a very elongated photoionized region that may
appear as a cm continuum ``radio jet.'' Later, the ionizing flux becomes strong
enough to begin to have an effect on the accretion flow, e.g., by
photoevaporation, although the presence of dust can significantly reduce
the effectiveness of this feedback (Tanaka et al., in prep.), in
contrast to the Population III case (McKee \& Tan 2008; Hosokawa et
al. 2011). If massive protostars do not have strong,
magnetocentrifugally-driven protostellar outflows, then their HII
region structures may be ``Accretion-Confined'' (Keto 2007; Peters et
al. 2010).

\subsection{Competitive Accretion and Protostellar Mergers}\label{S:CA}

Competitive Accretion (Bonnell et al. 2001; Wang et al. 2010; Smith et
al. 2011) involves protostars first forming from low-mass cores, with
masses typically set by the Bonnor-Ebert mass, i.e., $\ll
1\:M_\odot$ in $\Sigma_{\rm cl}\sim1\:{\rm g\:cm^{-2}}$
environments. The protostars then continue to accrete gas from the
clump and for intermediate-mass and high-mass stars this clump-fed,
Bondi-Hoyle mode of accretion plays the dominant role in setting the
final mass.

A significant difference between Competitive Accretion and Turbulent
Core Accretion is the former's much lower accretion rates to massive
protostars if the clump is not undergoing rapid global collapse, i.e.,
if the star formation efficiency per free-fall time, $\epsilon_{\rm
  ff}$, of the protocluster is $\lesssim 0.1$. This is seen in the
simulation of Wang et al. (2010), which includes outflow feedback that
helps stabilize the clump. The most massive star in the simulation
reaches 46.4~$M_\odot$ in 1~Myr, with an average accretion rate of only
$4.6\times 10^{-5}\:M_\odot\:{\rm yr}^{-1}$. In absence of this
feedback and in absence of $B$-fields that can help support
massive PSCs, rapid global collapse and fragmentation of the
clump is seen, leading to faster rates of Competitive Accretion
(Bonnell et al. 2001; Smith et al. 2011).

For massive protostars forming via Competitive Accretion, the
environment is required to be that near the center of a dense
protocluster, which is also crowded with lower-mass
protostars. Dynamical harassment of the protostar and its accretion
envelope and disk is much more severe. This will limit the size of
disks and lead to more rapid changes in their orientations, which will
also be reflected in the orientations of any associated protostellar
outflows. Simple bipolar outflows that maintain a fairly constant
orientation would not be expected in these models, also because of
presence of multiple overlapping, randomly-aligned outflows from the
surrounding lower-mass protostars. Such morphologies are also expected
to some extent for massive protostars forming via Turbulent Core
Accretion due to turbulence in the core, small-$N$ multiple formation
in the core, and neighboring protostars, so it is a question of the
degree to which relatively ordered morphologies are preserved in the
two scenarios, which still needs to be quantified in simulations.

Protostellar mergers (Bonnell et al. 1998), including via hardening of
binaries (Bonnell \& Bate 2005) have been proposed as a massive star
formation mechanism that operates in dense protocluster centers. For
collisional growth to be rapid compared to cluster formation or
massive stellar evolution timescales of $\sim 1$--10~Myr, requires
extreme stellar densities $\gtrsim 10^8\:{\rm pc^{-3}}$, equivalent to
$n_{\rm H}\gtrsim 3 \times 10^9\:{\rm cm}^{-3}$ (e.g., Moeckel \&
Clarke 2011). Efficient growth by mergers leads to runaway growth of
one or two extreme objects, rather than a smoothly filled upper
IMF. For these reasons, mergers are generally considered to be
unimportant in typical massive star-forming environments, although a
merger has been invoked to explain activity of the Orion KL protostar
(Bally \& Zinnecker 2005).

\section{Observational Constraints}

Here we discuss just a few recent examples of observations that test
massive star formation theories and compare to low-mass cases (see T14
for a more extensive review).

\subsection{Pre-Stellar Cores}\label{S:PSC}

One of the best studied ``low-mass'' PSCs is L1544 (Caselli \&
Ceccarelli 2012). However, this core actually has $\sim8\:M_\odot$,
and its slow, subsonic infall ($\lesssim$10\% of free-fall), suggests
$B$-fields play a significant role in its dynamics (Keto et al. 2015).

Searches for more massive PSCs in higher $\Sigma_{\rm cl}$
environments have focussed on IRDCs. Tan et al. (2013), following up a
sub-sample of the BT12 MIR extinction map peaks with {\it ALMA},
identified six cores via $\rm N_2D^+(3-2)$ emission, with the most
massive being C1-S with $\sim 60\:M_\odot$. The observed line-widths
are on average 80\% of that predicted by the fiducial Turbulent Core
model that assumes an Alfv\'en Mach number, $m_A=1$. However, for C1-S
the observed velocity dispersion is only 40\% of this level. Thus
virial equilibrium would require a stronger $B$-field of $\sim1$~mG
(i.e., $m_A=0.3$). This field strength, given the core density of
$n_{\rm H,c}\sim 6\times10^{5}\:{\rm cm}^{-3}$, also yields a magnetic
critical mass close to C1-S's observed mass, which could help explain
the core's limited fragmentation. Such $B$-field strengths are
similar to those predicted using the empirical relation $B\simeq
n_{\rm H}^{0.65}\:{\rm \mu G}$ (for $n_{\rm H}>300\:{\rm cm}^{-3}$)
(Crutcher et al. 2010) and are also similar to values inferred in some
massive protostars (\S\ref{S:proto}). The deuteration fraction,
$D_{\rm frac}^{\rm N2H+}\equiv[{\rm N_2D^+}]/[{\rm N_2H^+}]$, has been
measured in C1-S to be $0.2$--$0.7$ (Kong et al. 2015b),
several orders of magnitude greater than the cosmic [D]/[H]. By
comparison with chemodynamical models, Kong et al. (2015b) conclude it
is likely that C1-S is contracting at a relatively slow rate,
$\sim$1/10th of free-fall, so as to have had enough time to reach this
level of deuteration. This result needs confirmation by direct
measurement of infall speeds, but this is challenging given the
core is embedded in a much more massive and kinematically complex
clump environment.

C1-S provides physical and chemical evidence for the existence of a
massive, monolithic, centrally-concentrated, potentially-virialized
pre-stellar or early-stage core. G11.92-0.61-MM2 is another candidate
(Cyganowski et al. 2014), but based only on mm continuum emission \&
absence of line emission. The inferred mass of $\sim30\:M_\odot$
within $\sim 1000$~AU, i.e., $n_{\rm H}\gtrsim 10^{9}\:{\rm cm^{-3}}$,
yet without lines being seen from a lower density envelope, is
surprising.

\subsection{Protostellar Cores}\label{S:proto}

Nearby low-mass protostellar cores are being studied in unprecedented
detail with {\it ALMA}, including detection of rotating infall
envelopes, bipolar outflows, and potentially Keplerian disks on scales
$\lesssim 100$~AU (e.g., Codella et al. 2014). For the typically much
more distant massive protostars, there are clear examples of
collimated outflows (e.g., Beuther et al. 2002; Duarte-Cabral et
al. 2013) and ``rotating toroids'' (e.g., Beltr\'an et al. 2011;
S\'anchez-Monge et al. 2013). Infall has been detected in a number of
sources (e.g., Wyrowski et al. 2012). However, unambiguous detection
of disks remains challenging, which is not unexpected if diameters are
$\lesssim1000$~AU, i.e., $\lesssim0.5$'' at 2~kpc. Girart et
al. (2009) \& Q. Zhang et al. (2014) have inferred ``hour-glass''
morphologies of $B$-fields around some massive protostars, with field
strengths $\sim 1$~mG and concluded they play a dynamically important
role in massive star formation. Ionized, collimated outflows traced as
radio continuum ``jets'' have been seen in some massive protostars
(e.g., Gibb et al. 2003; Guzm\'an et al. 2014), although the relative
importance of shock- versus photo-ionization remains to be
established. The example of G35.20-0.74N contains all of the above
elements and detailed radiative transfer models based on Turbulent
Core Accretion have been fit to both its SED and resolved
multiwavelength images that probe outflow axis intensity asymmetries
to derive a protostellar mass of $m_*\sim 20$--$34\:M_\odot$ embedded
in a core with $M_c\simeq 240\:M_\odot$ in a clump with $\Sigma_{\rm
  cl}\simeq0.4$--$1\:{\rm g\:cm^{-2}}$ (Zhang et al. 2013b).

However, some massive protostellar cores appear much more disordered,
with the most famous example being the Orion KL ``hot core,'' which is
thought to be powered by a massive protostar detected as radio
``source I.'' There is an apparently ``explosive'' outflow from the
region, which has been interpreted as a signature of a protostellar
merger (Bally \& Zinnecker 2005). From X-ray images, Rivilla et
al. (2013) inferred a protostellar density of $\sim10^6\:{\rm
  pc^{-3}}$ in this region, still $\sim100\times$ less than that
required for efficient mass growth by mergers (\S\ref{S:CA}). However,
a single merger event involving a large disk-aided capture
cross-section of $\sim100$~AU may be expected to occur in
$\lesssim10^5$~yr. An alternative interpretation of Orion KL proposes
that the explosive outflow is caused by tidal harassment of source I's
accretion disk, leading to enhanced accretion and thus outflow, by a
passing runaway star, the BN Object (Chatterjee \& Tan 2012). In this
case, the interaction can be regarded as a moderate perturbation of
the Core Accretion model. Future proper motion studies of source I and
BN will help distinguish these scenarios.

\subsection{Protocluster Clumps}

Fragmentation of clumps and filaments is under active study (e.g.,
Beuther et al. 2013; Zhang et al. 2015), but interpretation of the
detected structures is hampered by lack of observational probes of
their $B$-fields. As reviewed by T14, infall times, $t_{\rm
  infall}\equiv M_{\rm cl}/\dot{M}_{\rm infall}$, relative to local
free-fall time, $t_{\rm ff}$, have been measured in several clumps,
including IRDCs, with typical observed ratios of $\sim 10$, indicating
slow, quasi equilibrium collapse. This suggests a dynamically
important role for $B$-fields, consistent with recent observations of
two IRDCs by Pillai et al. (2015), and/or stabilization of collapse by
protostellar outflow feedback, as modeled by Nakamura \& Li
(2014). As discussed in \S\ref{S:physics}, this question of the
dynamical state of the star-forming clump environment, including
nature of injected turbulence and distribution of $B$-field strengths,
is crucial for its effect on fragmentation to the CMF and thus the
formation mechanism of intermediate- and high-mass stars.

The quasi equilibrium cluster formation model (Tan et al. 2006)
predicts age spreads of stars $\sim 10 t_{\rm ff}$ in systems that
achieve final star formation efficiencies of $\sim 50\%$, consistent
with Da Rio et al.'s (2014) analysis of the Orion Nebula Cluster
(ONC). It also implies there is significant time during star cluster
formation for additional gas supply via infall from the clump
surroundings and for dynamical evolution of the stellar population
leading to mass segregation, so present-day locations of massive
stars, such as $\theta^1C$ in the ONC, are not likely to have been
where they formed. Improved observations of stellar kinematics (e.g.,
Foster et al. 2015; Cottaar et al. 2015) have the potential to test
these theories.


\end{document}